\documentclass{aa} 

\usepackage{amsmath}
\usepackage{amssymb}
\usepackage{natbib}
\bibpunct{(}{)}{;}{a}{}{,}
\usepackage{graphicx}
\usepackage{txfonts}
\usepackage[english]{babel}
\usepackage{hyperref}
\usepackage{units}
\hypersetup{colorlinks=true, linkcolor=blue, citecolor=blue, urlcolor=blue}
\usepackage{graphicx}
\usepackage{txfonts}
\begin{document} 

   \title{Surprising detection of an equatorial dust lane on the AGB star IRC+10216}

   \titlerunning{Equatorial dust lane on IRC +10216}

\author{
	S.~V.~Jeffers\inst{1,2}\thanks{The first two authors contributed equally to the work contained in this paper}
	\and
	M.~Min\inst{3,2\star}
	\and 
	L.~B.~F.~M.~Waters\inst{4,3}
	\and
	H.~Canovas\inst{5}
	\and
	O.~R.~Pols\inst{6}
	\and
	M.~Rodenhuis\inst{2}
	\and
	M.~de~Juan~Ovelar\inst{2,7}
	\and
	C.~U.~Keller\inst{2}
	\and
	L.~Decin\inst{8}
	\thanks{Based on observations made with the William Herschel Telescope operated on the island of La Palma by the Isaac Newton Group in the Spanish Observatorio del Roque de los Muchachos of the Instituto de Astrofísica de Canarias}}

\offprints{S.~V. Jeffers, \email{Jeffers@astro.physik.uni-goettingen.de}}

\institute{Institut f\"{u}r Astrophysik, Georg-August-Universit\"{a}t G\"{o}ttingen, Friedrich-Hund-Platz 1, 37077 G\"{o}ttingen, Germany
\and
Leiden Observatory, P.O. Box 9513, NL-2300 RA Leiden, The Netherlands 
\and
Sterrenkundig Instituut Anton Pannekoek, Universiteit van Amsterdam, Postbus 94249, 1090 GE Amsterdam, The Netherlands 
\and
SRON, Netherlands Institute for Space Research, 3584 CA Utrecht, The Netherlands 
\and
Departamento de Fisica y Astronomia, Universidad de Valparaiso, Valpariso, Chile
\and
Dept. of Astrophysics/IMAPP, Radbound Universiteit Nijmegen, Heijendaalseweg 135, 6525 AJ Nijmegen, The Netherlands 
\and
Astrophysics Research Institute, Liverpool John Moores University,146 Brownlow Hill, Liverpool L3 5RF, UK
\and
Instituut voor Sterrenkunde, Celestijnenlaan 200 D, B - 3001 Heverlee (Leuven), Belgium
}

   \date{\today}

 
  \abstract
   {}
   {Understanding the formation of planetary nebulae remains elusive because in the preceding asymtotic giant branch (AGB) phase these stars are heavily enshrouded in an optically thick dusty envelope. }
   {To further understand the morphology of the circumstellar environments of AGB stars we observe the closest carbon-rich AGB star IRC+10216 in scattered light.  }
   {When imaged in scattered light at optical wavelengths, IRC+10216 surprisingly shows a narrow equatorial density enhancement, in contrast to the large-scale spherical rings that have been imaged much further out.  We use radiative transfer models to interpret this structure in terms of two models: firstly, an equatorial density enhancement, commonly observed in the more evolved post-AGB stars, and secondly, in terms of a dust rings model, where a local enhancement of mass-loss creates a spiral ring as the star rotates.  }
{We conclude that both models can be used to reproduce the dark lane in the scattered light images, which is caused by an equatorially density enhancement formed by dense dust rather than a bipolar outflow as previously thought. We are unable to place constraints on the formation of the equatorial density enhancement by a binary system.} 


   \maketitle
%

\section{Introduction}

Low-to-intermediate mass stars evolve into large, luminous and cool giants at the end of their lives.  Their sudden death is caused by intense mass loss removing the star's outer envelope and fuel required for nuclear burning, leading to the formation of a planetary nebula.  In contrast to the highly asymmetrical shapes of planetary nebulae, the circumstellar environments of their immediate progenitor stars on the asymptotic giant branch (AGB) appear to be spherically symmetric.  Explaining how and when planetary nebulae are shaped remains elusive because of the difficulty in directly imaging the morphology of the preceding high mass loss phase, as these stars are heavily enshrouded in an optically thick dusty envelope. 

To further understand the morphology of the circumstellar environments of AGB stars, we observed the closest carbon-rich AGB star, IRC+10216 located at 123pc \citep{Groenewegen2012} which is also known as CW Leonis.  We apply the technique of polarimetric differential imaging using ExPo \citep[the Extreme Polarimeter][]{Rodenhuis2012} at the 4.2m William Herschel Telescope on La Palma.   Observing IRC+10216 in polarised light is advantageous as starlight scattered by a star's circumstellar material becomes linearly polarised and can be used to resolve stellar circumstellar environments in much greater detail than by intensity measurements alone \citep{Jeffers2013,Min2013,Canovas2012,Jeffers2012}.  The advantage of using imaging polarimetry is that it removes direct starlight and scattered light from the diffuse radiation field, both of which are unpolarised, leaving only the starlight scattered by its immediate circumstellar material.  The paper is organised as follows: in Section 2 we describe the observations and data analysis, in Section 3 we present the image of IRC+10216, which we interpret using radiative transfer models in Section 4.  In Section 5, we discuss the implications of our results.

\begin{figure*} 
\begin{center}
\includegraphics[scale=0.85]{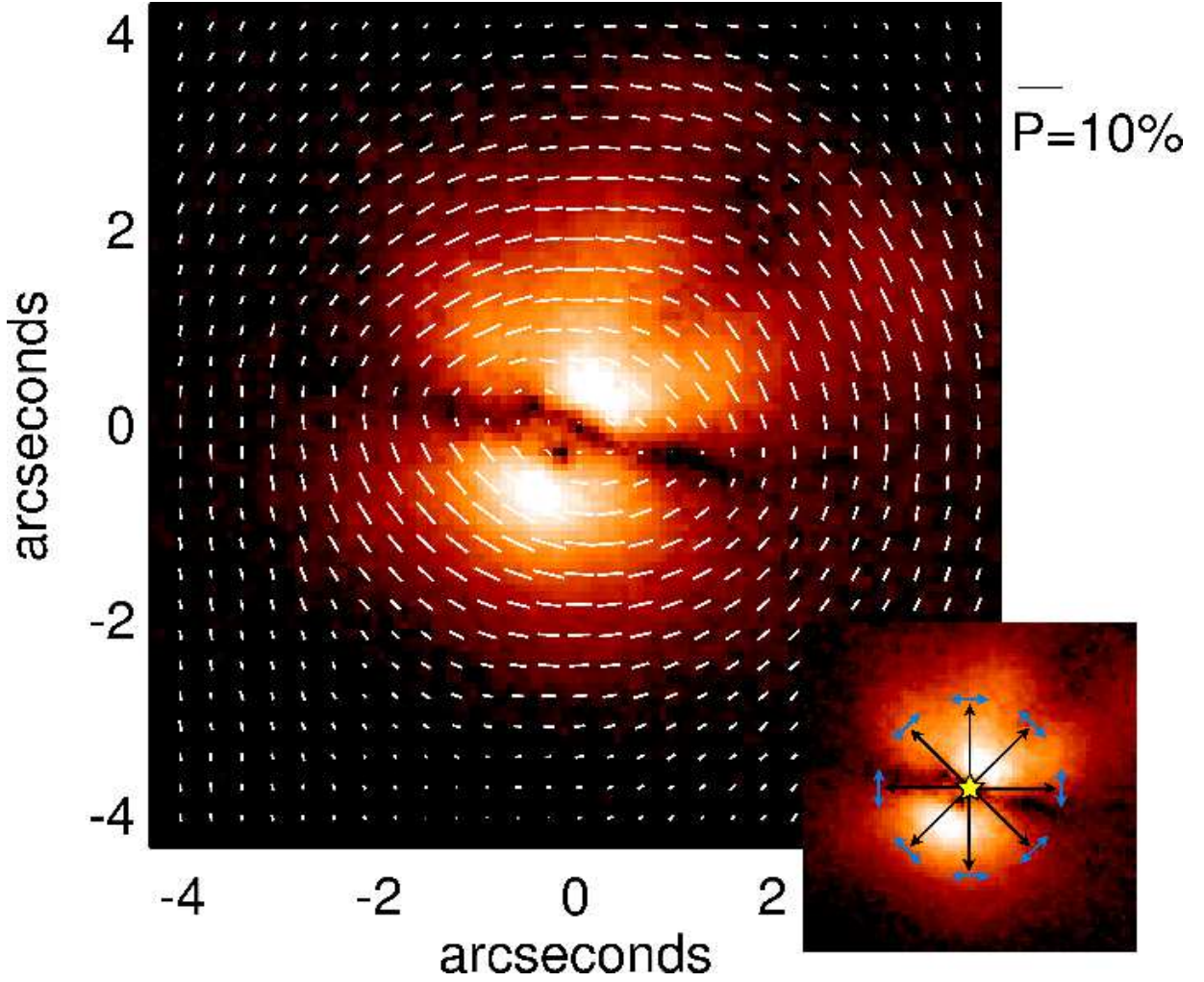}
\caption{ExPo image of the circumstellar environment of IRC+10216 in polarised intensity (P$_l$) observed at optical wavelengths 500-900nm.  The intensity scale is logarithmic and the length of the vectors represents a lower limit to the true polarisation degree.  North is up and east is to the left.  The inset panel shows how unpolarised starlight is scattered by dust in IRC+10216's circumstellar environment, resulting in linearly polarised light with a direction that is perpendicular to the incident light ray (in the reference frame of the observer).} 
\end{center} 
\protect\label{f-expoim} 
\end{figure*}

\section{ExPo observations and data analysis}

Observations of IRC+10216 were secured at the 4.2m William Herschel Telescope on La Palma using the dual-beam EXtreme Imaging Polarimeter, ExPo \citep{Rodenhuis2012}.  A ferro-electric liquid crystal (FLC) polarisation modulator is used to swap the polarisation states of the orthogonally polarised beams to minimise systematic effects.  The two beams are imaged simultaneously on an EMCCD (electron multiplier CCD \citep{Smith2004}) detector.  This results in four different images per FLC cycle: $A_\mathrm{left}$, $A_\mathrm{right}$, $B_\mathrm{left}$, $B_\mathrm{right}$, with $A$ and $B$ indicating the two orthogonal states.  Each frame is then flat-field corrected, dark subtracted, and cleaned of cosmic rays.   Individual frames are aligned with a sub-pixel precision with a template point-spread function.  The data reduction procedure is explained by \cite{Canovas2011}.  Images in linear polarisation are obtained from several thousand individual exposures (each having an exposure time of 0.2 s) that are combined by applying a double-difference approach \citep{kuhn2001,Hinkley2009}:

\begin{equation}
P'_I=0.5(\Delta A-\Delta B)=0.5((A_\mathrm{left}-A_\mathrm{right})-(B_\mathrm{left}-B_\mathrm{right}))
\end{equation}

and where the total intensity image is defined as:

\begin{equation}
I=0.5(A_\mathrm{left}+A_\mathrm{right}+B_\mathrm{left}+B_\mathrm{right}).
\end{equation}

At a given orientation of the FLC modulator, the instrument will be sensitive to polarisation along two orthogonal axes.  Observations are repeated at FLC angles of 0$^\circ$, 22.5$^\circ$, 45$^\circ$, and 67.5$^\circ$.  Finally, the reduced images are calibrated using the method of \cite{Rodenhuis2012} to produce Stokes Q and U images.  The  polarised intensity is defined as: $P_I$=$\sqrt{Q^2+U^2}$ and the degree of polarisation as $P=PI/I$.  The polarisation angle, which defines the orientation of the polarisation plane, is $P_\theta =0.5 arctan (U/Q)$.  The instrumental polarisation is removed in the data analysis by assuming that the central star is unpolarised.  We extensively tested the procedure of removing the instrumental polarisation to ensure that the dark lane is not an artifact of the data reduction pipeline.  More details can be found in \cite{Canovas2011} and Jeffers et al.(2013).

\section{Polarimetric image}


The circumstellar environment of IRC+10216 is shown in Figure~\ref{f-expoim} as imaged in linearly polarised light. The image shows a clear north-south axi-symmetric circumstellar shell, extending out to 4" from the central star in the east-west direction, and clear direct evidence of a dark lane.  The inset panel shows how unpolarised starlight is scattered by dust in IRC+10216's circumstellar environment and shows that the symmetric pattern of the polarisation vectors around the centre of the ExPo image indicates that the light produced by the star is scattered by its circumstellar environment.   The dark lane across the image indicates that the polarised flux in this area is significantly reduced compared to the rest of IRC+10216's circumstellar shell. This decrease in polarised flux can be explained by light travelling through an optically thick density enhancement of dust in front of a light source, which is forward scattered and consequently has lower polarisation.  This region of low polarisation creates a clear dark lane in our images. 

\section{Interpretative models}

To interpret the morphology of the polarisation image we use the radiative transfer code MCMax \citep{Min2009} to fit the spectral energy distribution (SED) of IRC+10216 and simultaneously compare the resulting output with the linearly polarised ExPo image.  This lifts many of the degeneracies normally associated with fitting the SED alone.  

\subsection{Model setup}

MCMax computes the full radiative transfer through the envelope including polarisation, where the particle composition, size, and shape are key parameters. For the composition of the dust particles, we use a mixture of amorphous carbon (88\%), SiC (10\%), and MgS (2\%).  These values are typical values for a carbonaceous dusty outflow \citep[for e.g.][]{Srinivasan2010,Hony2002}.  By fitting the SED and comparing the result with the scattered light images, we can constrain the albedo and density structure of the grains.  Since we do not perform a detailed mineralogical fit, the precise composition is not important for the large-scale morphology of the dust component.   

The refractive indices are taken from \cite{Preibisch1993,Pitman2008,Begemann1994}, for the amorphous carbon, SiC, and MgS respectively. For the size distribution of the grains we use $n(a) da \propto a^{-3.5}$, ranging from 0.005 micron up to 0.25 micron \citep{MathisRumplNordsieck1977}. We use irregularly shaped DHS (distribution of hollow spheres) grains, with the irregularity parameter $f_\mathrm{max}$=0.8, to avoid unrealistic polarisation behaviour present in the computations of perfect homogeneous spherical particles \citep{Min2005}.  The stellar parameters that we use are: effective temperature, $T_\mathrm{eff}$ = 2010 K and luminosity $L$ = 7790 L$_{\odot}$ \citep{Groenewegen2012}.

\subsection{Density distribution}

To understand the density distribution of IRC+10216, and, in particular, the dark dust lane, we use two different models.  The first is based on a classical model of the more evolved post-AGB stars with a thick equatorial torus (the `torus' model).  The second model is inspired by the mass ejection model of \cite{LeBertre1988} and is empirically postulated from our observations.  In this model intense local mass loss results in ring-like structures drifting away from the star (the `rings' model). 

\subsubsection{The torus model}

In the torus model the density structure of the circumstellar envelope of IRC+10216 is based on a well established model that explains the density distributions of the more evolved post-AGB stars \citep{Ueta2003}, where an equatorially enhanced dust torus is a common feature.  This density distribution has the following form:

\begin{eqnarray}
\rho (R,\theta) & = & \rho_0\left(\frac{R}{R_0}\right)^{\displaystyle -B\left\{1+C\sin^F\theta\left[\frac{e^{-(R/R_{sw})^D}}{e^{-(R_{min}/R_{sw})^D}}\right]\right\}} \nonumber \\
 && \qquad \times\left\{1+A(1-\cos\theta)^F\left[\frac{e^{-(R/R_{sw})^E}}{e^{-(R_{min}/R_{sw})^E}}\right]\right\},
\end{eqnarray}

where $R$ is the distance to the centre of the star, $R_\mathrm{sw}$ is the radius of the
super-wind region, $\theta$ is the latitude measured from the pole,
$\rho_0$ is the density at the inner edge of the shell, $R_0$ is the radius of the inner edge, and $A-F$ are fitting parameters. For $R \gg R_\mathrm{sw}$, the mass loss is spherically symmetric, while for $R \ll R_\mathrm{sw}$, the mass loss is equatorially enhanced.  There are two important quantitative parameters that can be derived from the model: (i) the density enhancement of the dust-torus compared to the circumstellar shell, and (ii) the outer radius of the dust torus, or super-wind radius, which physically represents the interface between the older spherical mass loss and the recent equatorial mass loss.  

\subsubsection{The rings model}


The rings model is based on the concept first presented by \cite{LeBertre1988}, where there are local enhancements in the mass loss rate that are for a short time similar to the ejection of dust puffs on R CrB stars \citep{Jeffers2012}.  Because of stellar rotation, the ejected mass moves away from the star and, as with the torus model, dust will form when the ejected mass reaches the dust condensation radius.  This results in density enhancement in the circumstellar matter for a given latitude band.  The ejected material has the same rotation axis as the star and can be shifted along the stellar rotation axis, however, we can only model complete rings because of the circular symmetry of our computational code.  The equations and parameters that define this model are empirically based on the features observed in the ExPo images, where we assume that one of the rings is currently forming across the equator of the star.  
  
In this model, the density distribution of the homogeneous, overall wind is equated by
\begin{equation}
\rho_\mathrm{owind}(R)=\rho_0 \left(\frac{R_0}{R}\right)^2,
\end{equation}
where $R$ is the distance from the central star and $\rho_0$ is the density at distance $R_0$.  The density distribution of a single ring $k$ is equated as follows
\begin{equation}
\rho_k(R,\theta)=\begin{cases}
\displaystyle \gamma\rho_0 \left(\frac{R_0}{R}\right)^2\frac{\left(1-|\sin(\theta-\theta_k)|\right)^6}{\sin\theta_k}, & R_k < R < (R_\mathrm{k}+R_\mathrm{ring}) \\
0,
\end{cases}
\end{equation}

where $\gamma$ is the density enhancement factor of the ring, $\theta$ is the angle with respect to the polar axis, $\theta_k$ is the angular location of the ring, and $R_\mathrm{ring}$ is the radial extent of the ring. Thus, an equatorial ring has $\theta_k=90^\circ$.
The total density becomes

\begin{equation}
\rho(R,\theta)=\rho_\mathrm{owind}(R)+\sum_k \rho_k(R,\theta)
\end{equation}

The most important parameters of this model are the density enhancement in the rings and the spatial extent of the rings radially and in latitude.  

\begin{figure*} 
\begin{center}
\centerline{\resizebox{0.95\hsize}{!}{\includegraphics{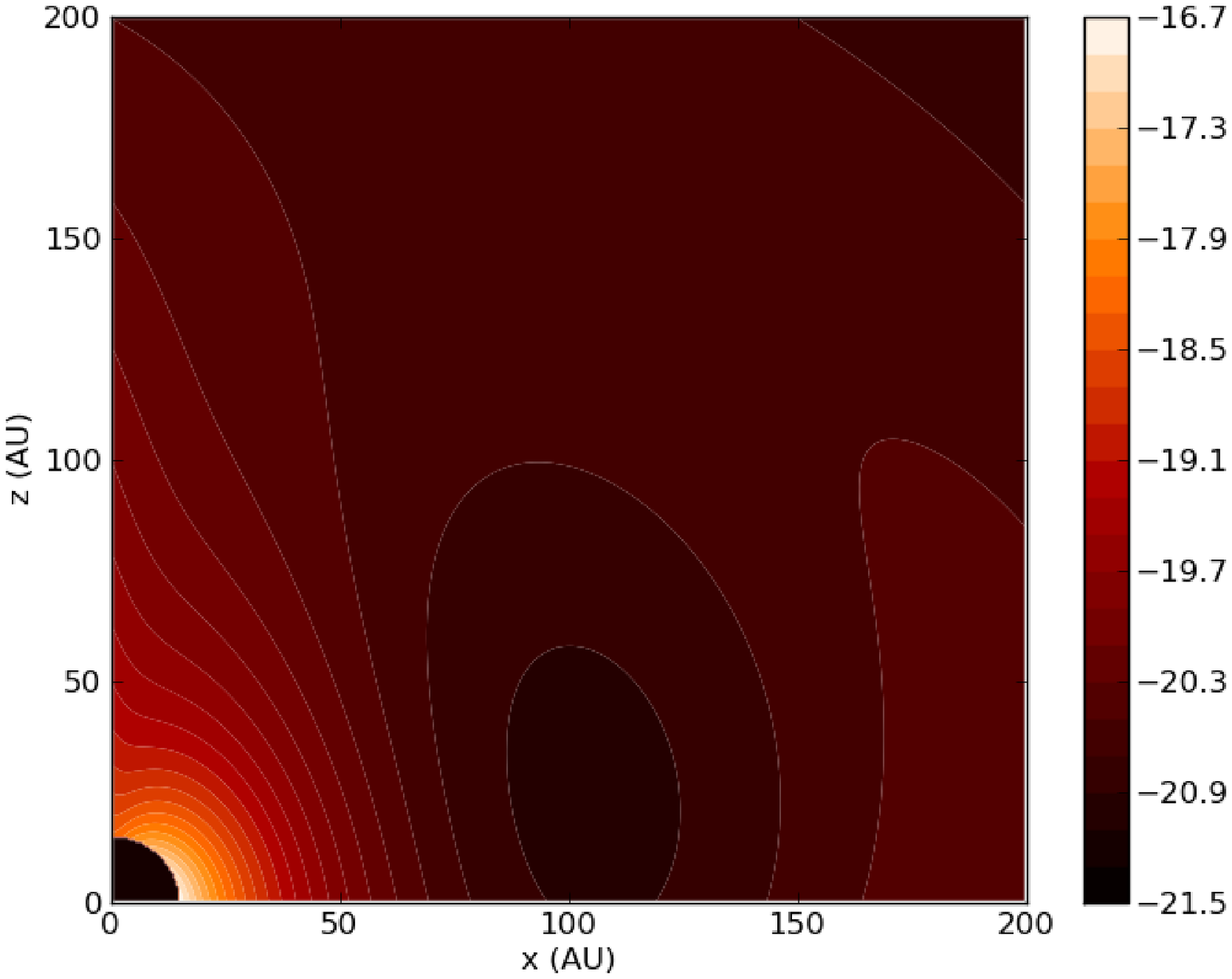}\includegraphics{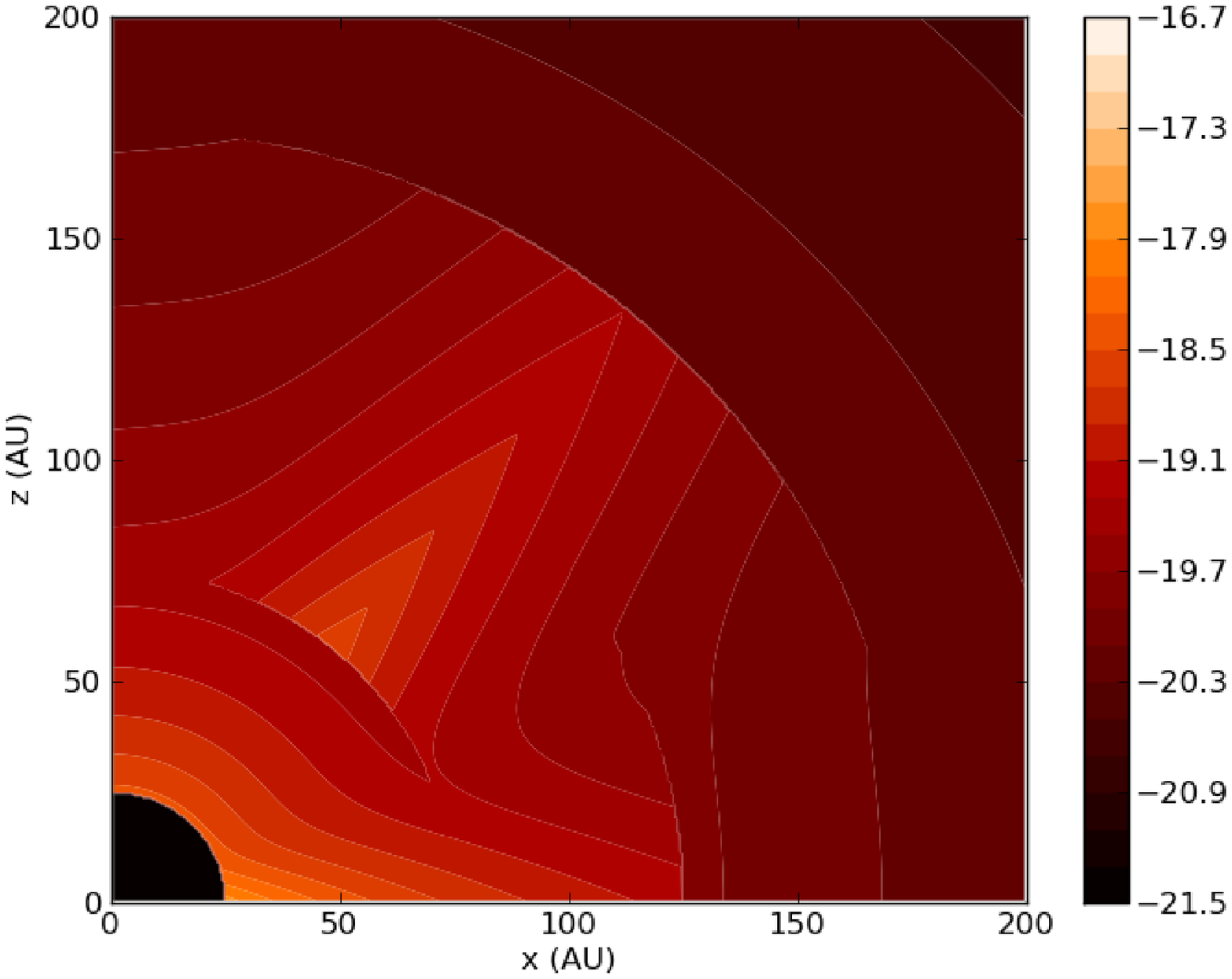}}}
\caption{The resulting dust density distribution of the circumstellar environment of IRC+10216 with the torus model (left) and the rings model (right).  The vertical axis is along the rotation axis of the star and the horizontal axis is in the plane of rotation.  The contour scale is logarithmic.}  
\protect\label{f-density} 
\end{center}
\end{figure*}

\begin{figure*} 
\begin{center}
\centerline{\resizebox{0.9\hsize}{!}{\includegraphics{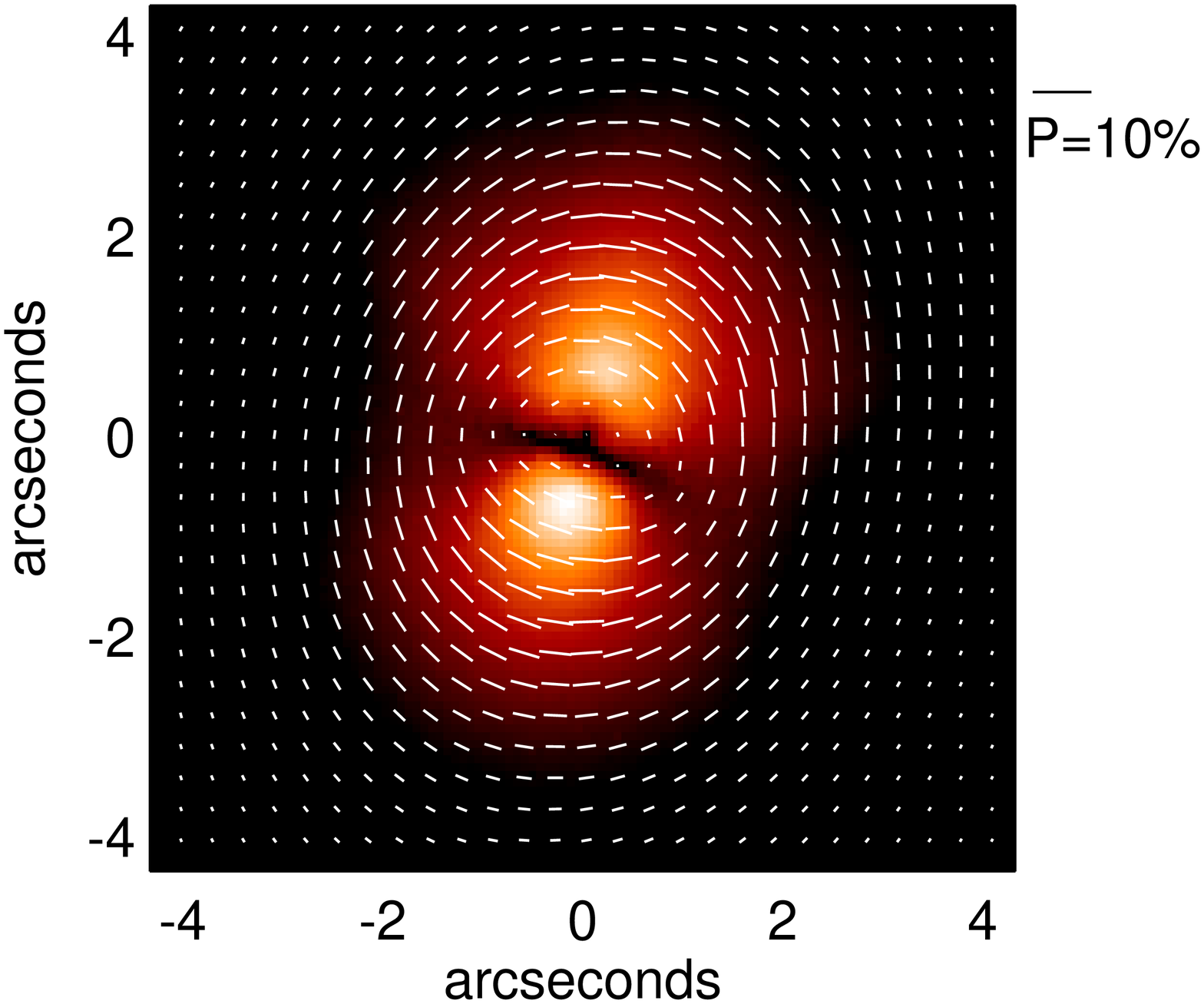}\includegraphics{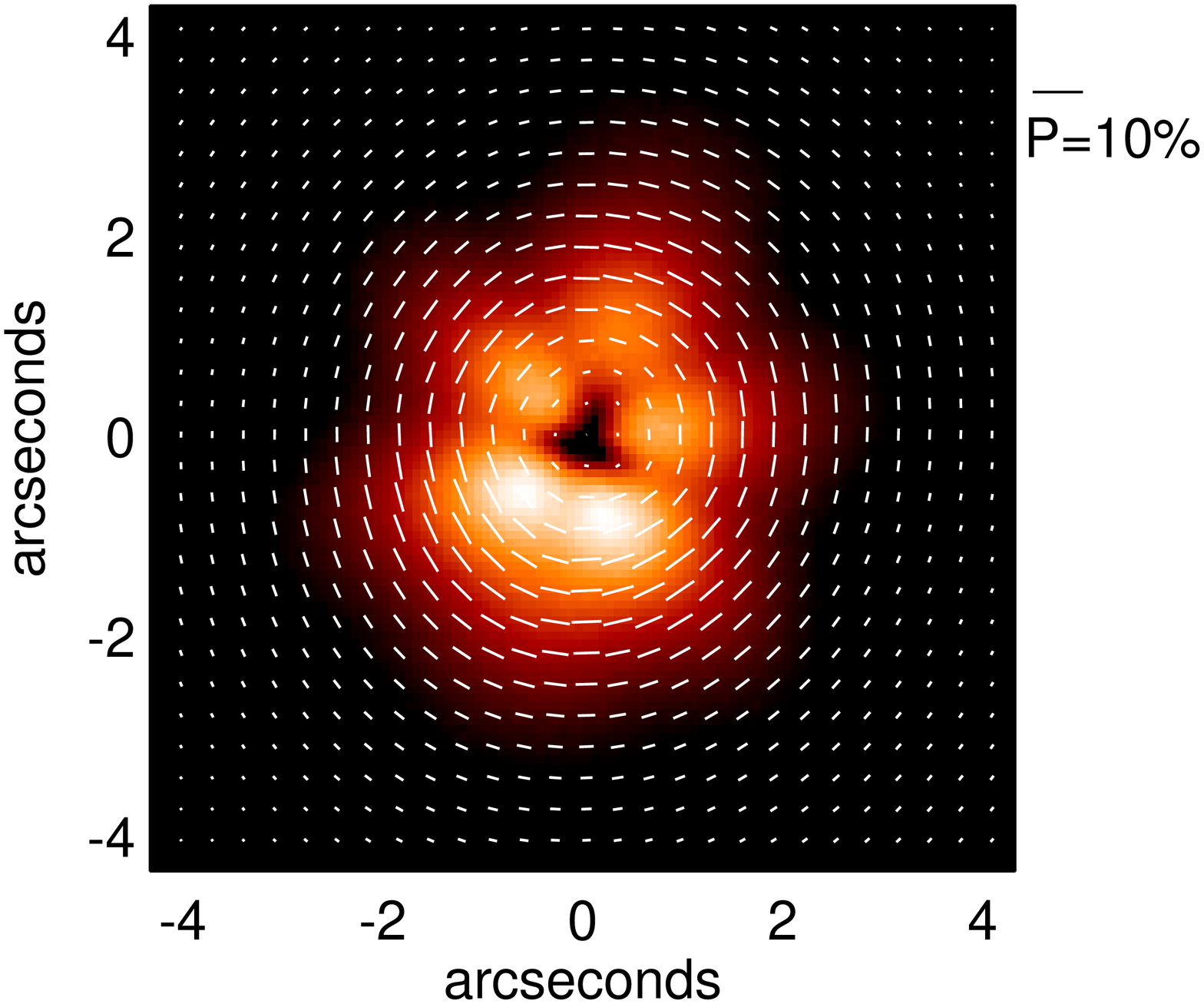}}}
\caption{Simulated image resulting from our interpretative model of IRC+10216 with the torus model (left) and the rings model (right) that includes seeing, instrument, and data reduction effects. }  
\protect\label{f-images} 
\end{center}
\end{figure*}

\subsection{Model results}
\label{sec:modelresults}

The derived parameters of the equatorial density enhancement were computed by fitting the SED using a genetic fitting algorithm and comparing the resulting image with the global properties of the ExPo images, which is consequently much more constraining than fitting the SED alone.  Since IRC+10216 is totally obscured by a thick dusty shell, the density distribution is continuous and it is not possible to distinguish separate contributions of the star, torus, and the outflow in the shape of the SED.  We focus on fitting the global shape of the SED with particular emphasis on the fit at optical and near infrared wavelengths where scattering is important.  The fit of the SED at longer wavelengths is used to check for consistency.  

To compare the results of our models with the linearly polarised image taken with ExPo, we convolve the resulting image with an instrumental point spread function (PSF) simulator that includes atmospheric seeing, instrument, and telescope effects. These models are then processed through the data reduction pipeline, as previously described, to account for any data reduction artifacts \citep[for more details see][]{Min2012}.  It is beyond the scope of this study to fit the ExPo image mathematically using an automated optimization routine. Instead, we examine the possible parameter space to capture the main characteristics of the model. The most important model parameters of the two models are presented in Table~\ref{tab:model_parameters}.



\subsection{Simulated images}

The simulated polarised intensity images of the two models are shown in Figure~\ref{f-images}.  Globally, they both reproduce the dark lane and the two bright (north-south) lobes seen in the ExPo image.  Both models also confirm that the dark lane seen in the ExPo images results from the central star being obscured by an optically thick dust torus, which is almost edge-on to the observer.  The dark lane occurs because starlight that is scattered towards the observer is forward scattered and unpolarised, resulting in a dark lane or shadow in the linearly polarised ExPo image. Below we discuss the two models used in this analysis to interpret the circumstellar environment of IRC+10216. The specific values of the parameters we used can be found in Tables~\ref{tab:general_parameters} and ~\ref{tab:model_parameters}.

\subsubsection{The torus model}

We find a good match to both the SED and the polarised light image using a density enhancement in the dust torus of a factor of approximately 30 compared to the poles of the circumstellar dust shell.  The optical depth in the visible is $\sim80$ along the line-of-sight and $\sim11$ in the polar direction. The outer radius $R_\mathrm{sw}$ is derived to be 150 AU to fit the ExPo observations, with the inner radius of the dust torus set to be at the dust condensation radius.  From this model of the density structure as a function of distance from the star, we can determine the average dust mass loss rate by integrating over a spherical surface surrounding the star. Assuming an outflow velocity of 10 km s$^{-1}$, we infer that the current dust mass loss rate, taken to be at the inner edge of the torus, is approximately 9 x 10$^{-8}$M$_\odot$ yr$^{-1}$.  This is a factor of 16 greater than the older spherical mass loss rate, which we determine to be 5.6 x 10$^{-9}$M$_\odot$ yr$^{-1}$ at a radius outside the super-wind radius.


The final density distribution of the circumstellar environment of IRC+10216 using the density distribution of the torus model is a radially compact, dense torus that obscures the star, with only the polar regions exposed.  The radial extent of the torus is approximately 30 AU.  Even though the dust torus is very compact, it creates a large shadow when observed in scattered light.  The density distribution of the torus model is shown in the left-hand panel of Figure~\ref{f-density}.

\begin{table}[!tbp]
\caption{Values for the fixed model parameters for both interpretive models.   In both models, we use a black body for the star.}
\begin{center}
\begin{tabular}{llc}
\hline
\hline
Fixed Parameters				&	Symbol		&	Value \\
\hline
Effective temperature	&	$T_\mathrm{eff}$	&	$2010\,$K \\
Radius of the star		&	$R_\star$			&	$730\,R_{\sun}$ \\
Luminosity of the star	&	$L_\star$			&	$7785\,L_{\sun}$\\
Distance to the star		&	$D$				&	$123\,$pc \\
Minimum grain size		&	$a_\mathrm{min}$			&	$0.005\,\mu$m\\
Maximum grain size		&	$a_\mathrm{max}$		&	$0.25\,\mu$m\\
Size distribution power law	&	$p$				&	$3.5$\\
Irregularity parameter	&	$f_\mathrm{max}$	&	$0.8$ \\
\hline
\hline
\end{tabular}
\end{center}
\label{tab:general_parameters}
\end{table}

\begin{figure} 
\includegraphics[scale=0.35]{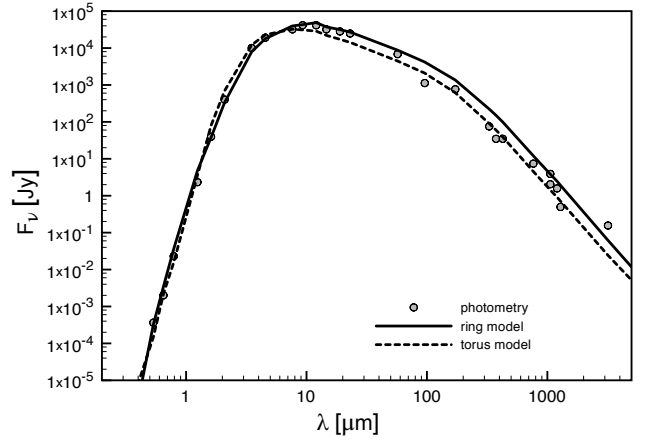}
\caption{The fitted SED of IRC+10216 using the model parameters described in Section~\ref{sec:modelresults}.  The data points are taken from \cite{bagnulo1995}.}  
\protect\label{f-sed} 
\end{figure}

\subsubsection{The rings model}

In this model we have derived a constant dust mass loss rate of 10$^{-8}$ M$_\odot$ yr$^{-1}$ with episodic, local increases in mass loss rate, which is determined empirically to fit the ExPo observations. We find that at the location of the eruption, which is fixed at the equator, the dust mass loss has to be increased by about a factor of 4 to match the ExPo observations. However, in the overall dust mass loss rate (i.e. integrated over a spherical surface surrounding the star), this is only a factor of 2. We find that a good match to both the SED and the ExPo image is obtained using a radial extent of the rings of $\sim100\,$AU and an latitudinal extent of $\sim30^\circ$. Assuming an outflow velocity of 10\,km/s this translates to a timescale of $\sim50$ years meaning that it is still possible to form a ring despite the very slow rotation of the star.  

To explain the bipolar structure observed in the ExPo image, it is necessary that the closest ring to the star is currently obscuring the star at equatorial latitudes.  Additionally, to reproduce the prominent three-finger structure, it is necessary to fix the location of a second ring at 40$^\circ$.  Because our models are azimuthally and north-south symmetric, a hint of the three-finger structure also appears in the southern lobe of the model image. 

The final density distribution of the rings model is much vertically flatter and extends out to approximately 100 AU from the star.  Additionally, there is the presence of the second ring at polar latitudes.  The density distribution of the rings model is shown in the right-hand panel of Figure~\ref{f-density}.

\begin{table}[!tbp]
\caption{Values for the best-fit model parameters for the torus model and the rings model}
\begin{tabular}{llc}
\hline
\hline
Fitted Parameter for torus model	&	Symbol		&	Value \\
\hline
Inclination angle		&	$i$				&	$87^\circ$\\
Inner radius			&	$R_0$			&	$15\,$AU \\
Superwind radius		&	$R_\mathrm{sw}$			&	$150\,$AU \\
Dust density at inner edge	&	$\rho_0$			&	$5.6\cdot10^{-19}\,$g/cm$^3$\\
Density enhancement		&	$A$				&	31\\
parameters (A-F)		&	$B$				&	2.0\\
					&	$C$				&	2.0\\
					&	$D$				&	2.8\\
					&	$E$				&	2.8\\
					&	$F$				&	1.0\\
\hline
Fitted Parameter rings model	\\ 
\hline
Inclination angle		&	$i$				&	$80^\circ$\\
Inner radius			&	$R_0$			&	$25\,$AU \\
Dust density at inner edge	&	$\rho_0$			&	$3.6\cdot10^{-19}\,$g/cm$^3$\\
Density enhancement 	&	$\gamma$		&	$3.7$\\
Radial extend of the ring	&	$R_\mathrm{ring}$	&	$100\,$AU \\
Closest ring			&	$\theta_1$		&	$90^\circ$ \\
					&	$R_1$			&	$25\,$AU \\
Second ring			&	$\theta_2$		&	$40^\circ$ \\
					&	$R_2$			&	$75\,$AU \\
\hline
\hline
\end{tabular}
\label{tab:model_parameters}
\end{table}

\section{Discussion} 

In this paper, we have presented the first direct images of an equatorial density enhancement around IRC+10216 using imaging polarimetry at optical wavelengths and at medium spatial scales.  
  Compared to previous studies, the dark lane of the optically thick density enhancement is clearly seen in our scattered light images because it is a scattering rather than an absorption effect.  

\subsection{Previous Observations}

IRC+10216 has been observed from very large (100s of arcseconds) to very small (milli-acrseconds) spatial scales, mostly at infrared wavelengths.  Many of these studies have found evidence for axi-symmetric structure \cite{Mauron2000,Skinner1998,Murakawa2002,Leao2006} and have only indirectly inferred the possible existence of an equatorial density enhancement of gas and dust in the inner regions of the wind.  However, so far with direct imaging, conclusive evidence for such a density enhancement has been elusive because observations were either done at infrared wavelengths, using regular intensity imaging, or with coronographs that obscure the inner regions of the circumstellar shell.  

Examples of previous observations where a dust disk has been inferred include the Hubble Space Telescope (HST) images of \citep{Skinner1998}, which are on a similar spatial scale as the ExPo images, and the images of \cite{Tuthill2000} on a milli-acrseconds spatial scale and the polarimetric images of \citep{Murakawa2005}, also on a sub-arcsecond scale.  The global morphology of the near-infrared K-band images of \cite{Tuthill2000} are consistent with the later images of \cite{Leao2006}, with an opaque region in the eastward side of the image, which \cite{Tuthill2000} postulate to be a dark lane in the east-west direction, consistent with our ExPo images.  However, this conjecture is not backed up by radiative transfer models \citep{Tuthill2005}.  The slightly larger-scale H-band near-infrared images of \cite{Murakawa2005} show a similar global morphology as \cite{Tuthill2005} and \cite{Leao2006}, but with the opaque region shifted to the southeast, and consequently the inferred dust torus is in the southeast-northwest direction. The differences with our results is that the opaque region is not symmetric across the image, and does not clearly indicate the presence of a dark lane across the image.  Other images at sub-arcsec spatial scales and taken at infrared wavelengths at J, H, and K wavelength bands\citep{Leao2006,Weigelt2002,Osterbart2000} do not find evidence of an equatorial density enhancement of dust, despite showing the same global morphology as \cite{Tuthill2000}.  Apparently at the small scale of these images, the environment is more complex.  It is not possible to compare the images directly with the ExPo images as with all polarimetric imaging studies, we exercise caution in interpreting small-scale polarisation features that fall within the central unpolarised PSF.  Similarly, we infer the central star to be at the central point of the polarisation vectors, implying that the central star is obscured by the equatorial density enhancement.


IRC+10216 is classified as a Mira variable undergoing stellar pulsations, resulting in a change of temperature, and radius which cause a large variation in luminosity.  We observed IRC+10216 serendipitously at the time of its photometric maximum as phased by the period of 638-d and the ephemeris of \citep{LeBertre1988} in January 2010 (at 11.49) and it was still visible in May 2010 (at 11.71 cycles).  However, in a recent observing run in November 2013 with the same instrumental setup, it was not possible to observe IRC+10216 as it was too faint.  This epoch of observation was 13.72 cycles after the epoch of \cite{LeBertre1988}, and almost exactly two full cycles after the May 2010 observations, making the non-detection quite surprising.  Photometric data points from the AAVSO (American Association of Variable Star Observers) show the magnitude of IRC+10216 in November 2013 to be approximately 18 at visual wavelengths, and much lower than our observations in May 2010 given that our minimum observable magnitude in May 2010 was approximately 13.0.

We note that the asymmetrical shape of the scattered light images provides a natural explanation for the discovery of water in the circumstellar environment of IRC +10216 \citep{Decin2010}.  Such a non-homogeneous structure is required for the interstellar UV photons to penetrate and start the photo-ionisation process of the main molecules near the star, which will finally react to form water.  Both of our models show that the axi-symmetric structure is caused by equatorial mass loss and is not a bi-polar outflow as previously described \citep[e.g.][]{Murakawa2002}.

\subsection{`Torus' model}

An obvious first explanation for the dark dust lane observed in the ExPo images is that we are witnessing the onset of the post-AGB phase where optically thick dust torii are quite common features.  As shown in our `torus' model it is possible to fit the SED and main features of the ExPo image quite well using the post-AGB dust torus model of \cite{Ueta2003}.   If this is indeed the case, then we are fortunate to witness the onset of axi-symmetric mass loss, starting as recently as 75 years ago, if we assume a typical outflow velocity for IRC+10216 of 10 km/s.  Additional recent N-band (8-13 micron) observations show a change in the shape of the SiC spectral feature at 11 micron and a reduction in the continuum at 13 microns, which is interpreted as a change in the mass loss history over the last 20 years \citep{Males2012}. 

During the formation of the dust torus, we infer from our model that the dust mass loss increased by a factor of 16. Remarkably, this is similar to the density enhancements that were previously derived for the older spherical arcs \citep{Decin2011}.  An inconsistency in this interpretation is that there has been an unexplained change in the mass loss geometry, while the mass loss rate itself has stayed quite constant. 


\subsection{Binary Interaction}

Such a dramatic change in mass loss behaviour seems hard to understand in both single star and binary models.  On arrival on the AGB, single stars are expected to rotate extremely slowly due to the enormous expansion of the envelope during previous stellar evolution phases and the strong loss of angular momentum in the star's stellar wind. This slow rotation will produce roughly spherical mass loss, i.e. without any preferred direction, as has been observed around IRC +10216 on large spatial scales \citep{Decin2011}.  However, to produce equatorially enhanced mass loss it is necessary that the star has been substantially spun up or that it is undergoing interaction with a binary companion.  Possible mechanisms include envelope spin-up from a potentially fast-rotating stellar core via thermal pulses \citep{Garcia-Segura1999}, envelope contraction if IRC +10216 is already starting to contract at the end of its AGB phase \citep{Heger1998}, or the interaction with a low mass stellar or planetary companion \citep[][and references therein]{Mayer2013}.  This interaction could cause density enhancements in the winds of AGB stars \citep{Kim2012}, which result in large-scale spiral density waves.

The latest observations with the HERSCHEL/PACS satellite show a binary frequency of  25.4\% in a sample of AGB stars \citep{Cox2012} and binarity is considered the most likely explanation for the spiral structures observed in the circumstellar material of the less evolved AGB star R Scl \citep{Maercker2012}.  Binarity has also been suggested to cause the arc-like structure seen in Herschel/PACS \citep{Decin2011} and HST data, but currently no binary models exist that can explain the irregular spacing of the arcs, or that they are often slightly geometrically offset from the central star. 

The question arises whether the close-in  density enhancement discussed in this paper could be associated with the series of arcs further out by means of a binary model. Since we view a significant density enhancement in the line of sight to the central star, this implies that in the context of a binary model the binary orbital plane must be close to edge-on.  \cite{Kim2012} show that in that case the spiral density structure shows significant deviations from ``spherical".  A more realistic model for the scattered and polarised intensity distribution of such binary induced density enhancements than presented in this paper is needed to answer the question whether our ExPo image could be due to binary interaction. 


\subsection{`Rings' model}

An alternative explanation of our observations is that IRC+10216 is undergoing a localised mass loss event based on the concept of \cite{LeBertre1988}.  In the current ejection event, the orientation of the ejected mass is such that it causes a dust stream or density enhancement that is seen as the dark lane observed in our polarised light images.  It is not implausible that IRC +10216 can have R CrB-like ejections of localised material \citep{Jeffers2012}, which would result in a clumpy structure rather than an homogeneous disk of material.   The presence of such a clumpy structure has been previously observed in the sub-arcsec observations of \cite{Leao2006,Weigelt2002,Osterbart2000}.

In this interpretation, such mass loss events together with the rotation of the star create rings that drift away from the star and could potentially cause the arc-like structures seen at large distances \citep{Decin2011}.  Previously, such ejections could have taken place at random latitudes on the (slowly rotating) star, causing the production of expanding arcs in random directions.  Because of the low optical depth at large distances, limb brightening could cause these rings to appear as incomplete arcs.   In the current event, the dust stream will continue to drift radially at equatorial latitudes.


\subsection{Comparison with the ExPo images}

The fundamental features of the ExPo image that need to be explained by the models are (1) the narrow dark lane that extends across the image in polarised light, and (2) the large-scale geometry of the image: e.g. the shape of the northern lobe, with three fingers, and the more spherical southern lobe and, (3) how well the model fits IRC+10216's SED.  An additional constraint from other large-scale images are the commonly observed large spherical rings.  We briefly summarise how each of these points can be used to ascertain how well each model interprets the circumstellar environment of IRC+10216.


\subsubsection{Narrow dark lane}

Both models can reproduce the equatorial dark lane that is visible in the ExPo images.  The torus model has been extensively used to interpret similar structures commonly observed around the more evolved post-AGB stars, however, this is the first time that the `rings' model has been presented using radiative transfer models in polarised light.  The `rings' model has the advantage that it could potentially also explain the clumpiness of the dust as observed in the small sub-arcsec scale images of \cite{Leao2006,Weigelt2002,Osterbart2000}.  

\subsubsection{Global geometry}

The global geometry of IRC+10216 shows a very distinctive shape in the form of two north-south spherical lobes with a smaller scale structure in the form of three fingers seen in the northern lobe.  Both models show that the axi-symmetric structure is caused by a dense dust torus of  equatorial mass loss and is not a bi-polar structure as previously described by \cite{Murakawa2002}.  

The three-finger shape of the northern lobe has previously been observed by \cite{Murakawa2002}, and it can also be discerned in the images of Skinner et al (1998).   The torus model can only produce three fingers in a very limited range of model parameters where a slight change in any of the input parameters yields only a two-finger structure. The fingers can be explained by scattered light emanating from the sides of IRC+10216's bipolar light-cone.  The fingers do not appear in the southern light-cone as IRC+10216 is inclined at approximately 10$^\circ$ from edge-on (with the northern light-cone facing towards us).  For the `rings' model, the three fingers can be reproduced when an second ring is fixed at a latitude of 40$^\circ$ and just north of the star.  This second ring would have been formed at an earlier epoch and has started drifting away from the star.  Because our models are azimuthally and north-south symmetric, the three-finger structure also appears in the southern lobe of the model image and is not exactly the same as the northern lobe as IRC+10216 is slightly tilted.

\subsubsection{SED fit}

Both models fit the SED to a satisfactory level especially at optical and near-infrared wavelengths where scattering is important.  At longer wavelengths, the torus model would appear to fit the SED marginally better, but given the uncertantities in the models, it is impossible to draw any definitive conclusions.  

\subsubsection{Large-scale rings}

The large-scale spherical rings, which are commonly observed in images of IRC+10216, are impossible to reproduce using the dust torus model.  The presence of these spherical rings on the large-scale images, and the dust torus on smaller scales, would imply that there was a sudden change in the mass loss geometry.  Given the derived current dust mass loss rates, which are computed by assuming a typical AGB outflow velocity of 10 km s$^{-1}$.  This would have only occurred over very recent timescales.  One advantage of the `rings' model is that it can potentially explain the large-scale rings without needing any sudden change in the mass loss geometry.  Even though the rings are elliptical in shape as we view IRC+10216 almost perpendicular to its rotation axis, they will appear as incomplete arcs as the edges of the rings are limb brightened.


\section{Conclusions}

Our observations of IRC+10216 in linear polarisation show for the first time the clear detection of an equatorial density enhancement around an AGB star.  We use two models of the density distribution of circumstellar material around IRC+10216 as input to our radiative transfer model, one following the classical torus model commonly observed on post-AGB stars, and a model where localised mass loss produces spherical rings.  This rings model has an advantage as it is empirically based, and future observations of AGB stars in polarised light will be able to confirm or refute the validity of this model.  However, the classical dust torus model has the advantage that it is well established and has been successfully applied to many systems.  While we are unable to determine the precise mechanism of equatorially enhanced mass loss, we can definitively conclude from our interpretive models that the dark lane is formed by dense dust lane at equatorial latitudes and not a bi-polar outflow as previously thought.

\begin{acknowledgements}

S.V.J. acknowledges research funding by the Deutsche Forschungsgemeinschaft (DFG) under grant SFB 963/1, project A16, H.C.C. acknowledges support from Millenium Science Initiative, Chilean Ministry of Economy, Nucleus P10-022-F.

\end{acknowledgements}

\bibliographystyle{aa}
\bibliography{iau_journals,cwleo}

\begin{thebibliography}{40}
\expandafter\ifx\csname natexlab\endcsname\relax\def\natexlab#1{#1}\fi

\bibitem[{{Bagnulo} {et~al.}(1995){Bagnulo}, {Doyle}, \&
  {Griffin}}]{bagnulo1995}
{Bagnulo}, S., {Doyle}, J.~G., \& {Griffin}, I.~P. 1995, A\&A, 301, 501

\bibitem[{{Begemann} {et~al.}(1994){Begemann}, {Dorschner}, {Henning},
  {Mutschke}, \& {Thamm}}]{Begemann1994}
{Begemann}, B., {Dorschner}, J., {Henning}, T., {Mutschke}, H., \& {Thamm}, E.
  1994, ApJ, 423, L71

\bibitem[{{Canovas} {et~al.}(2012){Canovas}, {Min}, {Jeffers}, {Rodenhuis}, \&
  {Keller}}]{Canovas2012}
{Canovas}, H., {Min}, M., {Jeffers}, S.~V., {Rodenhuis}, M., \& {Keller}, C.~U.
  2012, \aap, 543, A70

\bibitem[{{Canovas} {et~al.}(2011){Canovas}, {Rodenhuis}, {Jeffers}, {Min}, \&
  {Keller}}]{Canovas2011}
{Canovas}, H., {Rodenhuis}, M., {Jeffers}, S.~V., {Min}, M., \& {Keller}, C.~U.
  2011, A\&A, 531, A102+

\bibitem[{{Cox} {et~al.}(2012){Cox}, {Kerschbaum}, {van Marle}, {Decin},
  {Ladjal}, {Mayer}, {Groenewegen}, {van Eck}, {Royer}, {Ottensamer}, {Ueta},
  {Jorissen}, {Mecina}, {Meliani}, {Luntzer}, {Blommaert}, {Posch},
  {Vandenbussche}, \& {Waelkens}}]{Cox2012}
{Cox}, N.~L.~J., {Kerschbaum}, F., {van Marle}, A.-J., {et~al.} 2012, A\&A,
  537, A35

\bibitem[{{Decin} {et~al.}(2010){Decin}, {Ag{\'u}ndez}, {Barlow}, {Daniel},
  {Cernicharo}, {Lombaert}, {De Beck}, {Royer}, {Vandenbussche}, {Wesson},
  {Polehampton}, {Blommaert}, {De Meester}, {Exter}, {Feuchtgruber}, {Gear},
  {Gomez}, {Groenewegen}, {Gu{\'e}lin}, {Hargrave}, {Huygen}, {Imhof},
  {Ivison}, {Jean}, {Kahane}, {Kerschbaum}, {Leeks}, {Lim}, {Matsuura},
  {Olofsson}, {Posch}, {Regibo}, {Savini}, {Sibthorpe}, {Swinyard}, {Yates}, \&
  {Waelkens}}]{Decin2010}
{Decin}, L., {Ag{\'u}ndez}, M., {Barlow}, M.~J., {et~al.} 2010, Nat, 467, 64

\bibitem[{{Decin} {et~al.}(2011){Decin}, {Royer}, {Cox}, {Vandenbussche},
  {Ottensamer}, {Blommaert}, {Groenewegen}, {Barlow}, {Lim}, {Kerschbaum},
  {Posch}, \& {Waelkens}}]{Decin2011}
{Decin}, L., {Royer}, P., {Cox}, N.~L.~J., {et~al.} 2011, A\&A, 534, A1

\bibitem[{{Garc{\'{\i}}a-Segura} {et~al.}(1999){Garc{\'{\i}}a-Segura},
  {Langer}, {R{\'o}{\.z}yczka}, \& {Franco}}]{Garcia-Segura1999}
{Garc{\'{\i}}a-Segura}, G., {Langer}, N., {R{\'o}{\.z}yczka}, M., \& {Franco},
  J. 1999, ApJ, 517, 767

\bibitem[{{Groenewegen} {et~al.}(2012){Groenewegen}, {Barlow}, {Blommaert},
  {Cernicharo}, {Decin}, {Gomez}, {Hargrave}, {Kerschbaum}, {Ladjal}, {Lim},
  {Matsuura}, {Olofsson}, {Sibthorpe}, {Swinyard}, {Ueta}, \&
  {Yates}}]{Groenewegen2012}
{Groenewegen}, M.~A.~T., {Barlow}, M.~J., {Blommaert}, J.~A.~D.~L., {et~al.}
  2012, A\&A, 543, L8

\bibitem[{{Heger} \& {Langer}(1998)}]{Heger1998}
{Heger}, A. \& {Langer}, N. 1998, A\&A, 334, 210

\bibitem[{{Hinkley} {et~al.}(2009){Hinkley}, {Oppenheimer}, {Soummer},
  {Brenner}, {Graham}, {Perrin}, {Sivaramakrishnan}, {Lloyd}, {Roberts}, \&
  {Kuhn}}]{Hinkley2009}
{Hinkley}, S., {Oppenheimer}, B.~R., {Soummer}, R., {et~al.} 2009, AJ, 701, 804

\bibitem[{{Hony} {et~al.}(2002){Hony}, {Waters}, \& {Tielens}}]{Hony2002}
{Hony}, S., {Waters}, L.~B.~F.~M., \& {Tielens}, A.~G.~G.~M. 2002, A\&A, 390,
  533

\bibitem[{{Jeffers} {et~al.}(2013){Jeffers}, {Min}, {Canovas}, {Rodenhuis}, \&
  {Keller}}]{Jeffers2013}
{Jeffers}, S.~V., {Min}, M., {Canovas}, H., {Rodenhuis}, M., \& {Keller}, C.~U.
  2013, ArXiv e-prints

\bibitem[{{Jeffers} {et~al.}(2012){Jeffers}, {Min}, {Waters}, {Canovas},
  {Rodenhuis}, {de Juan Ovelar}, {Chies-Santos}, \& {Keller}}]{Jeffers2012}
{Jeffers}, S.~V., {Min}, M., {Waters}, L.~B.~F.~M., {et~al.} 2012, A\&A, 539,
  A56

\bibitem[{{Kim} \& {Taam}(2012)}]{Kim2012}
{Kim}, H. \& {Taam}, R.~E. 2012, AJ, 759, 59

\bibitem[{{Kuhn} {et~al.}(2001){Kuhn}, {Potter}, \& {Parise}}]{kuhn2001}
{Kuhn}, J.~R., {Potter}, D., \& {Parise}, B. 2001, ApJ, 553, L189

\bibitem[{{Le{\~a}o} {et~al.}(2006){Le{\~a}o}, {de Laverny}, {M{\'e}karnia},
  {de Medeiros}, \& {Vandame}}]{Leao2006}
{Le{\~a}o}, I.~C., {de Laverny}, P., {M{\'e}karnia}, D., {de Medeiros}, J.~R.,
  \& {Vandame}, B. 2006, A\&A, 455, 187

\bibitem[{{Le Bertre}(1988)}]{LeBertre1988}
{Le Bertre}, T. 1988, A\&A, 203, 85

\bibitem[{{Maercker} {et~al.}(2012){Maercker}, {Mohamed}, {Vlemmings},
  {Ramstedt}, {Groenewegen}, {Humphreys}, {Kerschbaum}, {Lindqvist},
  {Olofsson}, {Paladini}, {Wittkowski}, {de Gregorio-Monsalvo}, \&
  {Nyman}}]{Maercker2012}
{Maercker}, M., {Mohamed}, S., {Vlemmings}, W.~H.~T., {et~al.} 2012, Nat, 490,
  232

\bibitem[{{Males} {et~al.}(2012){Males}, {Close}, {Skemer}, {Hinz}, {Hoffmann},
  \& {Marengo}}]{Males2012}
{Males}, J.~R., {Close}, L.~M., {Skemer}, A.~J., {et~al.} 2012, \apj, 744, 133

\bibitem[{{Mathis} {et~al.}(1977){Mathis}, {Rumpl}, \&
  {Nordsieck}}]{MathisRumplNordsieck1977}
{Mathis}, J.~S., {Rumpl}, W., \& {Nordsieck}, K.~H. 1977, ApJ, 217, 425

\bibitem[{{Mauron} \& {Huggins}(2000)}]{Mauron2000}
{Mauron}, N. \& {Huggins}, P.~J. 2000, A\&A, 359, 707

\bibitem[{{Mayer} {et~al.}(2013){Mayer}, {Jorissen}, {Kerschbaum},
  {Ottensamer}, {Nowotny}, {Cox}, {Aringer}, {Blommaert}, {Decin}, {van Eck},
  {Gail}, {Groenewegen}, {Kornfeld}, {Mecina}, {Posch}, {Vandenbussche}, \&
  {Waelkens}}]{Mayer2013}
{Mayer}, A., {Jorissen}, A., {Kerschbaum}, F., {et~al.} 2013, A\&A, 549, A69

\bibitem[{{Min} {et~al.}(2012){Min}, {Canovas}, {Mulders}, \&
  {Keller}}]{Min2012}
{Min}, M., {Canovas}, H., {Mulders}, G.~D., \& {Keller}, C.~U. 2012, A\&A, 537,
  A75

\bibitem[{{Min} {et~al.}(2009){Min}, {Dullemond}, {Dominik}, {de Koter}, \&
  {Hovenier}}]{Min2009}
{Min}, M., {Dullemond}, C.~P., {Dominik}, C., {de Koter}, A., \& {Hovenier},
  J.~W. 2009, A\&A, 497, 155

\bibitem[{{Min} {et~al.}(2005){Min}, {Hovenier}, \& {de Koter}}]{Min2005}
{Min}, M., {Hovenier}, J.~W., \& {de Koter}, A. 2005, A\&A, 432, 909

\bibitem[{{Min} {et~al.}(2013){Min}, {Jeffers}, {Canovas}, {Rodenhuis},
  {Keller}, \& {Waters}}]{Min2013}
{Min}, M., {Jeffers}, S.~V., {Canovas}, H., {et~al.} 2013, A\&A, 554, A15

\bibitem[{{Murakawa} {et~al.}(2005){Murakawa}, {Suto}, {Oya}, {Yates}, {Ueta},
  \& {Meixner}}]{Murakawa2005}
{Murakawa}, K., {Suto}, H., {Oya}, S., {et~al.} 2005, A\&A, 436, 601

\bibitem[{{Murakawa} {et~al.}(2002){Murakawa}, {Tamura}, {Suto}, {Itoh},
  {Hayashi}, {Oasa}, {Nakajima}, {Kaifu}, {Yates}, {Gledhill}, {Richards},
  {Hough}, {Kosugi}, \& {Usuda}}]{Murakawa2002}
{Murakawa}, K., {Tamura}, M., {Suto}, H., {et~al.} 2002, A\&A, 395, L9

\bibitem[{{Osterbart} {et~al.}(2000){Osterbart}, {Balega}, {Bl{\"o}cker},
  {Men'shchikov}, \& {Weigelt}}]{Osterbart2000}
{Osterbart}, R., {Balega}, Y.~Y., {Bl{\"o}cker}, T., {Men'shchikov}, A.~B., \&
  {Weigelt}, G. 2000, A\&A, 357, 169

\bibitem[{{Pitman} {et~al.}(2008){Pitman}, {Hofmeister}, {Corman}, \&
  {Speck}}]{Pitman2008}
{Pitman}, K.~M., {Hofmeister}, A.~M., {Corman}, A.~B., \& {Speck}, A.~K. 2008,
  A\&A, 483, 661

\bibitem[{{Preibisch} {et~al.}(1993){Preibisch}, {Ossenkopf}, {Yorke}, \&
  {Henning}}]{Preibisch1993}
{Preibisch}, T., {Ossenkopf}, V., {Yorke}, H.~W., \& {Henning}, T. 1993, A\&A,
  279, 577

\bibitem[{{Rodenhuis} {et~al.}(2012){Rodenhuis}, {Canovas}, {Jeffers}, {de Juan
  Ovelar}, {Min}, {Homs}, \& {Keller}}]{Rodenhuis2012}
{Rodenhuis}, M., {Canovas}, H., {Jeffers}, S.~V., {et~al.} 2012, in Society of
  Photo-Optical Instrumentation Engineers (SPIE) Conference Series, Vol. 8446,
  Society of Photo-Optical Instrumentation Engineers (SPIE) Conference Series

\bibitem[{{Skinner} {et~al.}(1998){Skinner}, {Meixner}, \&
  {Bobrowsky}}]{Skinner1998}
{Skinner}, C.~J., {Meixner}, M., \& {Bobrowsky}, M. 1998, MNRAS, 300, L29

\bibitem[{{Smith} {et~al.}(2004){Smith}, {Coates}, {Giltinan}, {Howard},
  {O'Connor}, {O'Driscoll}, {Hauser}, \& {Wagner}}]{Smith2004}
{Smith}, N., {Coates}, C., {Giltinan}, A., {et~al.} 2004, in Society of
  Photo-Optical Instrumentation Engineers (SPIE) Conference Series, Vol. 5499,
  Optical and Infrared Detectors for Astronomy, ed. J.~D. {Garnett} \& J.~W.
  {Beletic}, 162--172

\bibitem[{{Srinivasan} {et~al.}(2010){Srinivasan}, {Sargent}, {Matsuura},
  {Meixner}, {Kemper}, {Tielens}, {Volk}, {Speck}, {Woods}, {Gordon},
  {Marengo}, \& {Sloan}}]{Srinivasan2010}
{Srinivasan}, S., {Sargent}, B.~A., {Matsuura}, M., {et~al.} 2010, A\&A, 524,
  A49

\bibitem[{{Tuthill} {et~al.}(2005){Tuthill}, {Monnier}, \&
  {Danchi}}]{Tuthill2005}
{Tuthill}, P.~G., {Monnier}, J.~D., \& {Danchi}, W.~C. 2005, AJ, 624, 352

\bibitem[{{Tuthill} {et~al.}(2000){Tuthill}, {Monnier}, {Danchi}, \&
  {Lopez}}]{Tuthill2000}
{Tuthill}, P.~G., {Monnier}, J.~D., {Danchi}, W.~C., \& {Lopez}, B. 2000, AJ,
  543, 284

\bibitem[{{Ueta} \& {Meixner}(2003)}]{Ueta2003}
{Ueta}, T. \& {Meixner}, M. 2003, ApJ, 586, 1338

\bibitem[{{Weigelt} {et~al.}(2002){Weigelt}, {Balega}, {Bl{\"o}cker},
  {Hofmann}, {Men'shchikov}, \& {Winters}}]{Weigelt2002}
{Weigelt}, G., {Balega}, Y.~Y., {Bl{\"o}cker}, T., {et~al.} 2002, A\&A, 392,
  131

\end{thebibliography}

\end{document}